\documentclass[11pt,twoside]{article}

\usepackage{asp2006}
\usepackage{epsf}
\usepackage{psfig}
\usepackage{lscape}

\begin{document}
\title{The outer Galaxy: stellar populations and dark matter}   
\author{Gerry Gilmore}   
\affil{Institute of Astronomy, Cambridge, UK}    

\begin{abstract} 
The Galaxy's stellar populations are naturally classified into six
`types', of which five have been observed. These are the thin disk
(Pop~I in the historical scheme), a discrete thick disk (Pop~I.5), the
metal-rich bulge, which was not named in the Baade sequence, the rare
field halo (Pop~II), a population currently being accreted into the
very outer halo filed (Pop~Sgr?)and a hard to discover initial
enriching Pop~III. Each of these forms a group with astonishly tight
correlations between chemical element ratios and other parameters. It
is very hard to understand how the observed properties of any one of
these populations can be the sum of many discrete histories, except
for the minor continuing outer halo accretion. All these stellar
populations are embedded in dark-matter, and allow the properties of
dark matter to be measured on small scales. Intriguing and unexpected
consistencies in the properties of this dark matter are being
revealed.
\end{abstract}

\section{Stellar Population Types}

In the Milky Way we can identify five stellar populations with
properties that constrain the star formation history, the chemical
evolution history (flows, feedback..)  and the mass assembly history
of the Galaxy.  There may even be a sixth type, sometimes called Pop~III, as yet
undetected directly but postulated to provide the first chemical
enrichment
.
\begin{itemize}
\item{The thin disk, also known as Baade's Population I.  This is
composed of stars and gas on high angular momentum orbits, moving
about the center with close to the circular velocity, and thus with
only low amplitude random motions. Such a cold thin system presumably
formed by dissipational collapse of gas, in a potential that is
changing slowly, and conserved angular momentum to spin-up as it
collapsed (see Fall \& Efstathiou 1980; Mo, Mao \& White 1998).  The
origins of disks are however not clear: hierarchical merging models
predict significant angular momentum transport and generate disks that are too
small (Navarro \& Steinmetz 1997).  Appeal to some suitable process of
`feedback' can be implemented to prevent much of the angular momentum
losses from the proto-disk, but at the expense of delaying the
collapse to centrifugal equilibrium (e.g. Eke, Efstathiou \& Wright
2000) and thus predicting few old stars in disks, and no extended
high-redshift disks.  Late perturbations to the thin disk cannot be
too strong, or the disk will be destroyed (e.g.~Ostriker 1990). The
properties of stars in the thin disk are important tests of merging
histories and energetic dynamical processes.  The age and metallicity
distributions of the disk, well-defined only at the solar
neighbourhood, point to extended infall of metal-poor gas, and steady
star formation from a redshift of $\sim 1.5$ (e.g.~Binney et
al.~2000) to the present.}

\item{The thick disk - this was identified as a separate component
some 25 years ago (Gilmore \& Reid 1983).  The dominant population is
old, as old as the globular cluster 47~Tuc, $\sim 12$~Gyr, and of
intermediate metallicity in the mean, $[Fe/H] \sim -0.6$, with a
significant spread.  The chemical enrichment history revealed by the
pattern of element ratios is distinct from that of stars in the thin
disk (Bensby et al.~2007, this meeting).  A plausible origin for the
thick disk is a moderately violent dynamical event such as a minor
merger; the old mean age for the thick disk limits such events to have
occurred only long ago, an important constraint -- and a problem, if
found to be a typical result -- for CDM models. Thick disks are often
observed in resolved stars in other galaxies (e.g.~Mould 2005; Yoachim
\& Dalcanton~2005) but their properties remain to be robustly
determined.}

\item{The central bulge - interestingly, though probably most stars in
 the Universe are in spheroids, this was not in the classic Baade list
 of stellar populations. The dominant stellar population in the bulge
 of the Milky Way is old and metal-rich, with a broad spread in
 metallicities.  Elemental abundances are available for remarkably few
 stars, given the capabilities of current telescopes, but where
 available point to a fairly rapid enrichment, being dominated by
 products of Type II supernovae. This, together with the old age and
 high (phase-space) density, point to {\sl in situ} formation, in a
 `starburst', though a star formation rate of $\sim 10M_{\odot}
 yr^{-1}$ is all that is required, at high redshift. In some way, this
 is connected to the formation of the supermassive black hole at the
 Galactic Center.  The relationships between the overlapping `bulge',
 `bar' and the thin and thick disks in the inner kpc or so remain
 unclear.}

\item{The stellar halo, also known as Baade's Population II.  This is
a dominantly old and metal-poor component, with Type II dominated
element ratios, indicating a short duration of star formation in each
of the star-forming entities that created the halo. The negligible
scatter in the ratio of alpha elements to iron for stars with a range
in Galactocentric orbits of some tems of kpc is startling, and
suggests {\sl in situ} formation from a single well-mixed ISM over
scales of that size during halo formation.  Accretion to the dominant
inner Population II halo can only have occured at early times
(Unavane, Wyse \& Gilmore 1996).  The bulk of the Population II halo
may be connected to the stellar bulge; one can tie gas outflow from
halo star-forming regions, required to provide the low mean
metallicity, to gas inflow to the central regions to form the
bulge. The low angular momentum of proto-halo material means that it
will only come into centrifugal equilibrium after collapsing in radius
by a significant factor.  The predicted mass ratio of bulge to halo is
around a factor of ten, just as would be expected, and the specific
angular momentum distributions of stellar halo and bulge match (Wyse
\& Gilmore 1992; see Figure~1 here).}

\item{ The outer parts
of the halo, although containing only a small fraction of even the
rare halo stars, have a more complex structure and history (e.g the
`Field of Streams', Belokurov etal 2006). There are clear
indications of significant accretion, most dramatically due to
the Sagittarius dwarf (Ibata, Gilmore \& Irwin 1994)
which is currently populating the outer halo with mostly intermediate-age and
metal-rich members. It is quite unclear what the progenitor of Sgr
would have looked like a few Gyr ago.}

\item{Population III -- which we take to mean stars formed from
primordial gas,  precursers to `galaxy' formation.  Where
are the low-mass Pop III stars?  On-going searches for extremely low
metallicity stars in the Galactic halo have not found any strong
indications of a separate population (e.g.~Beers et al.~2005), but
have identified a few stars with extreme deficiencies in iron, and
relatively strong carbon (e.g.~Aoki et al.~2007).  The origins of this
abundance pattern are unclear. There is little observational evidence
in favour of significant variations in the stellar IMF for any of the
components discussed above, but there is theoretical prejudice
that primordial stars might form with a narrow range of masses, around $\sim
200$~M$_\odot$ (e.g. Bromm \& Larson 2004).  The supernovae from such
stars would provide elemental abundance patterns in the stars they
enrich that do not match those of the extremely metal-poor stars,
indicating that much remains to be learned.}

\item{Population Zero: the dark matter}

\end{itemize}

\begin{figure}[h!]
\begin{center}
\plotone{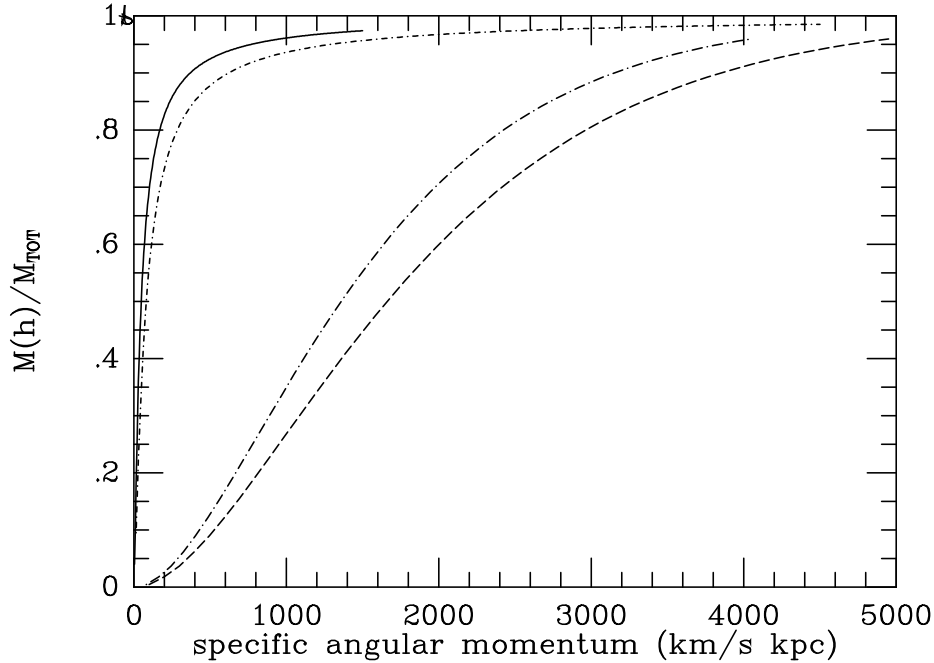}
\caption{Adapted from Wyse \& Gilmore 1992, their Figure~1.  Angular
momentum distributions of the bulge (solid curve), the stellar halo
(short-dashed/dotted curve), the thick disk (long-dashed/dotted curve)
and the thin disk (long-dashed curve).  The bulge and stellar halo
have similar distributions, as do the thick and thin disks. Does this
hold for external galaxies, pointing to fundamental relationships
between bulge and halo, and thick and thin disks? }
\end{center}
\end{figure}

\section{Dark Matter on small scales}

\begin{figure}[!h]
\begin{center}
\plotone{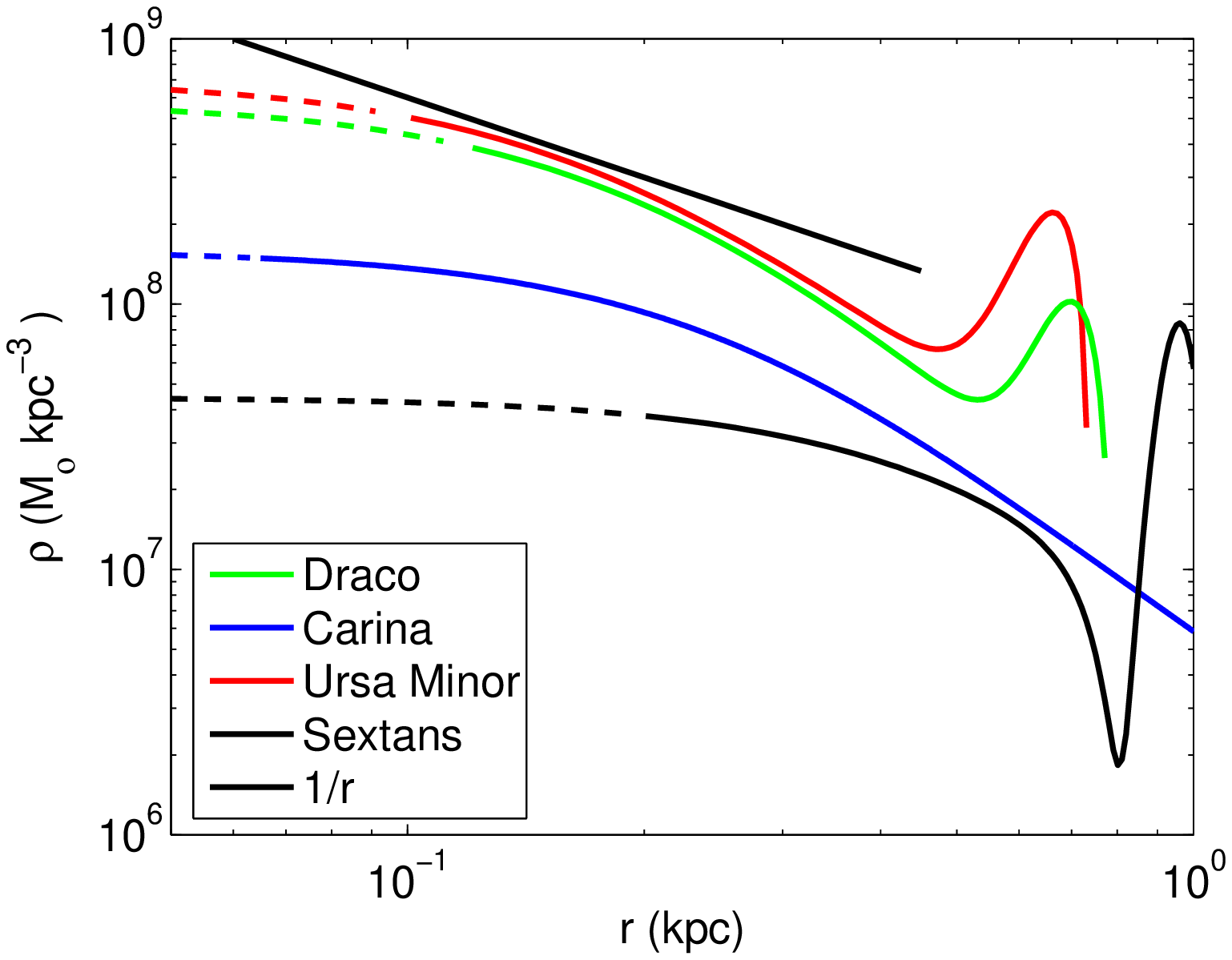}
\caption{Derived inner mass distributions from Jeans' eqn analyses for
four dSph galaxies. Also shown is a predicted $r^{-1}$ density
profile. The modelling is reliable in each case out to radii of log
(r)kpc$\sim0.5$. The unphysical behaviour at larger radii is explained
in the text. The general similarity of the four inner mass profiles is
striking.}
\end{center}
\end{figure}

The Milky Way satellite dwarf spheroidal (dSph) galaxies are the
smallest dark matter dominated systems in the universe.  Several groups have
underway dynamical studies of the dSph to quantify the shortest scale
lengths on which Dark Matter is distributed, the range of Dark Matter central
densities, and the density profile(s) of DM on small scales. An
updated overview of results will be presented in Gilmore et al (2007).

All mass distribution analyses based on the Jeans' equations - as most
are to date - involve an inherent degeneracy between mass and (stellar
tracer) orbital anisotropy. Nonetheless, the observed properties of
all the dSph studied to date, including their half-light radii, the
amplitude of their central velocity dispersions, and the flatness of
their velocity dispersion profiles, require, purely from observations,
a quite remarkable similarity among the least luminous galaxies, even
though they exist over several magnitudes in absolute magnitude.

Exploiting this similarity in both optical sizes and kinematic
properties, there is a simple consistency argument which links the
observed distribution of sizes of small galaxies, the clear
distinction in size and phase-space density between star clusters and
galaxies of the same total luminosity, and all the available dSph
galaxy dynamical analyses.  Our current results suggest some
surprising regularities: the central dark matter density profile is
typically cored, not cusped, with scale sizes never less than one
hundred pc; the central densities are typically $10-20$GeV/cc if the
mass is cored, and less than$\sim 1$TeV/cc even if the dark matter is
cusped. No galaxy is found with a dark mass halo less massive than
$\sim 10^7M_{\odot}$.

All the dSph analysed by all groups to date show very similar, and
surprisingly low, central dark matter mass densities, with a maximum
value of $\sim 0.5 M_{\odot} pc^{-3}$, equivalent to $\sim 20$GeV/cc,
if the mass profile really is cored. Interestingly, the rank ordering
of the central densities, though not robustly determined, seems in
inverse order to system total luminosity, with the least luminous
galaxies being the most dense. This is of the opposite sign to some
CDM predictions. The low maximum mass density is also intriguing,
given that some currently favoured dark matter candidate particles are
of Higgs scale, TeV mass - their corresponding volume density must be
very low indeed.

These consistencies were suspected, largely based on the results of
Figure~2 here, before the recent flurry of discoveries of several very
low luminosity dSph galaxies, and the availability of several new
kinematic studies. Interestingly, the validity of the conclusions is
becoming stronger as the sample and the dynamic range are improved,
suggesting some underlying general properties of dark matter on the
smallest scales are within observational reach (Gilmore etal 2007 -submitted)

We are discovering many more dSphs, which we and other groups are
analysing to test the generality of these results.

\acknowledgements 

This review is based on joint work with many collaborators, much also
summarised in Wyse and Gilmore 2006 (astroph-0604130), in Gilmore et al
2006 (astroph-0608528), and in a larger analysis presented in Gilmore
etal 2007 (ApJ - submitted).

\end{document}